\documentclass{article} 
\usepackage[final]{colm2026_conference}

\usepackage[utf8]{inputenc}
\usepackage{microtype}
\usepackage{hyperref}
\usepackage{url}
\usepackage{booktabs}
\usepackage{graphicx}
\usepackage{amsmath}
\usepackage{tikz}
\usetikzlibrary{arrows.meta, positioning}

\definecolor{darkblue}{rgb}{0, 0, 0.5}
\hypersetup{colorlinks=true, citecolor=darkblue, linkcolor=darkblue, urlcolor=darkblue}

\title{Context-Aware Prompt Injection Detection via Small Language
Models in RAG Pipelines: nokast-secureRAG}
\author{Abhishek Rai\\ Independent Researcher\\ \texttt{abhishek.rai085@gmail.com}}

\begin{document}

\ifcolmsubmission
\linenumbers
\fi

\maketitle
\lhead{Accepted at the Lifelong Agent Workshop, COLM 2026}

\begin{abstract}
Local Retrieval-Augmented Generation (RAG) systems let powerful assistants operate
over private data, but inherit a growing class of security risks---particularly
Indirect Prompt Injection (IPI) and retrieval poisoning, in which adversarial
instructions hidden in retrieved content are executed by the model as if
trusted~\citep{greshake,agentpoison,openragsoc}. This threat is no longer
hypothetical: a recent web-scale measurement validated $15.3$\,K indirect-injection
instances across $11.7$\,K pages, roughly $70\%$ of them hidden in non-rendered
HTML~\citep{ipi_wild}, and security teams report live agents coerced into payments
and data exfiltration~\citep{unit42_ipi,ipi_exfil}. Existing defenses are either
retrieval-centric (sanitization) or safety-centric guards trained on direct
jailbreaks rather than RAG context consistency~\citep{openragsoc,llamaguard}. We present
\emph{nokast-secureRAG}, a $\sim$0.5B-parameter Small Language Model (SLM) that acts
as a \emph{semantic firewall} between the retriever and the generator: it jointly
reads the user query $Q$ and the retrieved context $C$ and classifies the pair as
\textsf{safe}, \textsf{suspicious}, or \textsf{malicious-instruction}. The model is
produced by a \emph{teacher$\rightarrow$judge$\rightarrow$student} distillation
pipeline: a 35B teacher generates 5{,}233 realistic $(Q,C)$ records across four
attack families; an independent, larger 120B judge of a different model family
re-labels every record blind, yielding $95.0\%$ agreement; and the agreed data
fine-tunes the student via reasoning-first LoRA. On a held-out test set the tuned
SLM reaches $0.994$ detection recall at $0.026$ false-positive rate and $0.006$
attack-success-rate (detection-side proxy), versus $0.688$ recall for a regex
baseline, $0.888$ for a public prompt-injection classifier (ProtectAI
DeBERTa~\citep{protectai}), $0.437$ for Llama Guard 3 1B~\citep{llamaguard}, and $0.116$
for the same model untuned. On a dedicated \emph{judgment-flip} test---where an
identical sentence is benign in one context and malicious in another---the SLM labels
both halves correctly $75\%$ of the time versus $17.5\%$ for a keyword filter, $54\%$
for the public classifier, and $27\%$ for Llama Guard, directly demonstrating
context-aware detection; a fourth public baseline purpose-built for prompt-injection
detection (Prompt-Guard-86M~\citep{promptguard}) reaches high raw recall ($0.900$) but a $0.690$
false-positive rate from context-blind over-flagging of ordinary instructional text,
illustrating exactly the failure mode this paper's $(Q,C)$ reframing targets. Input
ablations confirm the attack signal lives in $C$ (masking $C$ collapses recall to
zero) while query--context consistency measurably reduces false positives beyond what
$C$ alone provides. Deployment-relevant
single-request latency is $293$--$404$\,ms depending on serving runtime---above this
paper's own sub-$100$\,ms target---with $37$\,ms amortized throughput under batching. We additionally evaluate
out-of-distribution on PoisonedRAG~\citep{poisonedrag}, an external knowledge-poisoning
benchmark: recall collapses to near-zero, a negative result we report in full because
PoisonedRAG's fluent fact-fabrication attacks fall outside this detector's
instruction-consistency threat model. All headline numbers remain in-distribution.
The model, data-generation pipeline, and evaluation harness are released open source.
\end{abstract}

\section{Introduction}
Local RAG combines a retriever with a language model to ground responses in external
knowledge while keeping data on user-controlled hardware~\citep{openragsoc}. Integrating
external sources, however, exposes the system to prompt injection and data
poisoning~\citep{greshake,agentpoison,openragsoc}. Unlike direct jailbreaks, \emph{indirect}
prompt injection embeds adversarial instructions in data that is later retrieved and
processed, blurring the boundary between ``data'' and ``instructions''~\citep{greshake}.Recent work also highlights the role of calibrated decision models in agent security \citep{Liu2026}.
Greshake et al. show such attacks can compromise real LLM-integrated applications;
Chan et al. show RAG memory/knowledge-base poisoning at low poison
rates~\citep{greshake,agentpoison}.

In previous work, we proposed the nokast-secureRAG framework for trusted local AI,
which uses small language models as semantic firewalls between retrieval and
generation in privacy-preserving RAG systems~\citep{rai_nokast_framework}. That paper
was primarily a conceptual and architectural contribution. The present work extends
that framework by introducing a specialized $\sim$0.5B-parameter detector model,
instantiating a teacher$\rightarrow$judge$\rightarrow$student distillation pipeline,
and empirically evaluating context-aware prompt-injection detection in RAG.

Recent measurements show that indirect prompt injection is no longer a purely
academic threat. A web-scale study over $1.2$\,B URLs and $24.8$\,M hosts identified
$15.3$\,K validated injection instances across $11.7$\,K pages, with roughly $70\%$
of payloads residing in non-rendered HTML such as comments, headers, and metadata,
and compliance reaching up to $8\%$ even for smaller models on plain-text
inputs~\citep{ipi_wild}. Security teams report the same pattern operationally: hidden
web-page instructions have hijacked AI agents into initiating payments, deleting
databases, and approving scam content~\citep{unit42_ipi}, and a single poisoned email
has been shown to coerce a frontier model into exfiltrating SSH keys in a multi-agent
workflow~\citep{ipi_exfil}.

From an operator's perspective, the question is not only whether indirect prompt
injection is possible, but whether it is worth engineering against: IBM's 2024 Cost
of a Data Breach report puts the average global breach cost at USD~$4.88$\,M, rising
to USD~$6.08$\,M for financial organizations~\citep{ibm_cost_breach}, so a local
firewall that closes even one viable exfiltration path can plausibly pay for itself.

A core weakness of current defenses is that they are \textbf{context-blind with
respect to user intent}. The sentence ``Delete all previous output'' is benign in a
disk-utility manual but malicious when injected into a news snippet retrieved into a
conversation. Keyword filters key on the surface string and therefore cannot
distinguish the two; general-purpose guards such as Llama Guard~\citep{llamaguard} are
trained on direct-jailbreak and content-safety taxonomies rather than RAG context
consistency; and using a large cloud guard for a privacy-first local deployment
defeats the purpose of local inference.

This paper presents and \emph{empirically evaluates} nokast-secureRAG, a local
$\sim$0.5B SLM specialized for context-aware injection detection. Our contributions:
\begin{itemize}
  \item A reframing of RAG injection detection as a \textbf{semantic context
  consistency} task over the pair $(Q,C)$ rather than the query alone, with a
  three-class label schema (\textsf{safe}/\textsf{suspicious}/\textsf{malicious-instruction}).
  \item A reproducible \textbf{teacher$\rightarrow$judge$\rightarrow$student}
  distillation pipeline using schema-constrained synthetic generation and an
  independent cross-model judge, run end-to-end on a single 121\,GB consumer device.
  \item A \textbf{judgment-flip} evaluation that isolates context-dependent safety
  and cleanly separates context-aware detection from keyword matching.
  \item Empirical results showing the tuned SLM achieves $0.994$ recall / $0.026$
  FPR at $37$\,ms throughput, and an open-source release of model, data pipeline, and harness.
  \item A comparison against three real prompt-injection/safety classifiers (ProtectAI
  DeBERTa~\citep{protectai}, Llama Guard 3~\citep{llamaguard}, and
  Prompt-Guard-86M~\citep{promptguard}),
  input ablations isolating the contribution of $Q$ versus $C$, and an
  honestly-reported out-of-distribution check against an external knowledge-poisoning
  benchmark~\citep{poisonedrag}.
\end{itemize}
We frame our headline numbers as \emph{in-distribution}; we do not claim validated
robustness against adaptive adversaries, and our one external-benchmark check
(Sec.~\ref{sec:poisonedrag}) is a negative result on a threat class outside this
detector's design scope, not a generalization claim.

\section{Background and Related Work}
We organize the literature into five threads: retrieval-augmented generation,
prompt-injection and jailbreak attacks, RAG-specific poisoning, benchmarks and
defenses, and the efficient-adaptation techniques our method builds on.

\subsection{Retrieval-Augmented Generation}
Retrieval-augmented generation (RAG) couples a parametric language model with a
non-parametric retriever, conditioning generation on documents fetched from an
external corpus~\citep{lewis}. RAG improves factuality and domain alignment and
underpins most local ``chat-with-your-data'' assistants, but it widens the trust
boundary: any content that can be retrieved can influence generation. The
OpenRAG-Soc harness standardizes end-to-end RAG evaluation from ingestion through
retrieval to answer generation~\citep{openragsoc}, which we adopt conceptually as the
deployment setting for our defense.

\subsection{Prompt Injection and Jailbreaks}
Adversarial control of LLM behavior spans \emph{direct} and \emph{indirect} attacks.
Direct attacks manipulate the user-visible prompt: Perez and Ribeiro show simple
``ignore previous instructions'' goal-hijacking and prompt-leaking~\citep{perez};
Wallace et al. find input-agnostic \emph{universal adversarial triggers}~\citep{wallace};
and Zou et al. construct transferable adversarial suffixes via greedy coordinate
gradient search that defeat alignment~\citep{gcg}. \emph{Indirect} prompt injection
(IPI) moves the payload into data the model later ingests: Greshake et al. formalize
IPI for LLM-integrated applications, demonstrating remote control, data exfiltration,
and denial of service via injected content with no direct user prompt~\citep{greshake};
Liu et al. systematize prompt-injection attacks against deployed LLM-integrated
applications~\citep{liu}. Beyond case
studies, Liu et al.\ formalize prompt injection and systematically benchmark five
attack families against ten defenses across multiple models and tasks, finding that
many ad-hoc filters and instruction-based defenses offer only limited robustness
under stronger attacks~\citep{promptinj_bench}. Our work is complementary: we focus on
a local, specialized defense layer for the retrieval-mediated subset of prompt
injection.

\subsection{RAG and Knowledge-Base Poisoning}
A complementary threat corrupts the retrieval corpus itself. Chan et al. (AgentPoison)
poison the long-term memory and knowledge bases of RAG/agentic systems, achieving high
attack success at low poison rates while preserving benign behavior~\citep{agentpoison}.
Zou et al. (PoisonedRAG) inject a small number of crafted passages so that target
queries retrieve attacker text and produce attacker-chosen answers~\citep{poisonedrag}.
These works motivate two of our attack families (\textsf{poisoning} and
\textsf{context-hijack}) and underscore that the defense must reason about content the
retriever ranks highly, not merely obvious keyword triggers.

\subsection{Benchmarks and Defenses}
Benchmarks increasingly stress-test IPI in realistic carriers. The OpenRAG-Soc
harness of Guo and Wei embeds injections in hidden HTML, alt text, and accessibility
attributes and ships deployable mitigations such as sanitization, Unicode
normalization, and attribution-gated answering~\citep{openragsoc}; Yi et al. (BIPIA) benchmark indirect injection
and evaluate black- and white-box defenses~\citep{bipia}. On the guard side, Llama
Guard is an LLM input--output safeguard trained on general safety
taxonomies~\citep{llamaguard}, and prompt-level defenses (quote-and-cite,
retrieval-aware critique, ``no new instructions from context'') attenuate risk but
still rely on the base model implicitly separating data from
instructions~\citep{openragsoc}. Complementary work benchmarks RAG security directly:
SafeRAG introduces a RAG-specific security dataset spanning silver-noise,
inter-context-conflict, soft-ad, and white-DoS attacks and shows that representative
RAG components remain vulnerable even to obvious attacks despite retrievers, filters,
and advanced LLMs~\citep{saferag}; a parallel benchmark of poisoning attacks against
RAG evaluates thirteen attack methods and seven defenses across diverse architectures
and finds them broadly susceptible~\citep{rag_poison_benchmark}. In parallel, a
guardrail ecosystem has emerged: Guardrails AI provides an open-source validator
framework for input/output guards~\citep{guardrails_ai}, and NVIDIA NeMo Guardrails
adds programmable topic, dialog, retrieval, and safety rails around LLM
applications~\citep{nemo_guardrails}. GuardBench standardizes guardrail-model
evaluation across $40$ datasets and finds that general-purpose instruction-following
models can match specialized guards on standard safety metrics~\citep{guardbench}.
Crucially, guardrails are themselves context-fragile: inserting benign retrieved
documents flips Llama-Guard safety judgments in roughly $8$--$11\%$ of
cases~\citep{rag_guardrails_unsafe}, which motivates a RAG-tailored component that
reasons over the query--context pair rather than over each text in isolation. These
efforts largely target either ingestion mechanics or generic safety; none provides a
small, RAG-specialized model that explicitly verifies query--context consistency
between retrieval and generation.

\subsection{Efficient Adaptation, Distillation, and Reasoning}
Our student is produced with techniques from three lines of work. Knowledge
distillation transfers the behavior of large teacher models into a compact
student~\citep{hinton}; Low-Rank Adaptation (LoRA) fine-tunes only small injected
low-rank matrices, leaving base weights frozen and yielding tiny, portable
adapters~\citep{lora}; and chain-of-thought supervision encourages a model to produce
intermediate reasoning before a final answer~\citep{cot}, which motivates our
reasoning-first label format. Combining these lets a 0.5B model inherit
teacher/judge-level judgment at a fraction of the inference cost.

\subsection{Positioning}
nokast-secureRAG is a complementary, \emph{local} defense layer that explicitly reasons
about the alignment between user intent and retrieved context before that context
reaches the generator---distinct from retrieval-centric
sanitization~\citep{openragsoc}, generic safety guards~\citep{llamaguard}, guardrail
frameworks~\citep{guardrails_ai,nemo_guardrails}, RAG security
benchmarks~\citep{saferag,rag_poison_benchmark}, and attack/poisoning
studies~\citep{greshake,agentpoison,poisonedrag} in being an \emph{implemented,
empirically evaluated} detector with a dedicated context-flip test.
Table~\ref{tab:compare} summarizes this positioning.

\begin{table}[t]
\centering
\caption{Qualitative positioning of nokast-secureRAG against representative guards,
guardrail frameworks, and RAG security benchmarks (citations in \S2.4). ``Reasons
over $(Q,C)$'' marks whether the component conditions its verdict on the user query
\emph{and} the retrieved context jointly; ``RAG-specialized'' marks whether it is
designed for the retrieval-mediated threat specifically; ``Local/small'' reflects
design intent (Llama Guard is not built as a RAG-local guard), not the specific
checkpoint size benchmarked in \S6, where B5 uses the 1B Llama Guard 3 checkpoint
locally for comparison.}
\label{tab:compare}
\resizebox{\textwidth}{!}{%
\begin{tabular}{lccccl}
\toprule
System & Local/small & Reasons over $(Q,C)$ & RAG-specialized & Open & Kind \\
\midrule
\textbf{nokast-secureRAG (ours)} & \textbf{Yes} & \textbf{Yes} & \textbf{Yes} &
\textbf{Yes} & detector \\
Llama Guard        & No  & No      & No      & Yes & guard model \\
Guardrails AI       & N/A & No      & Partial & Yes & framework \\
NeMo Guardrails     & N/A & Partial & Partial & Yes & framework \\
SafeRAG             & N/A & N/A     & Yes     & Yes & benchmark \\
\bottomrule
\end{tabular}%
}
\end{table}

\section{Threat Model and Problem Statement}
We consider a local RAG pipeline in which retrieved chunks are fused with the user
query and passed to a generator. The adversary controls some retrievable content
(web pages, documents, memory entries) and embeds instructions intended to redirect
the assistant: overriding system/user instructions, exfiltrating data, or redefining
the assistant's role. We assume the user query itself is benign; the attack lives in
the \emph{context}.

The defense must (i) detect instruction-like content in $C$ that conflicts with the
intent of $Q$, (ii) avoid over-flagging benign instructional text (utility), and
(iii) operate within local hardware budgets (16\,GB-class VRAM, interactive latency)
without a cloud guard. We refer to context-driven inconsistency in classification---a
phrase being safe in one context and unsafe in another---as a \emph{judgment flip},
and treat correct handling of flips as the defining capability of a context-aware
detector.

\section{Method}
\subsection{Semantic context consistency model}
The detector receives the structured input $\text{Input}=[\,Q\,]\,\Vert\,[\,C\,]$ and
emits a label in $\{\textsf{safe},\textsf{suspicious},\textsf{malicious-instruction}\}$
preceded by a short reasoning trace. Because the model reads $Q$ and $C$ jointly, it
can condition the verdict on whether instruction-like content in $C$ is consistent
with the user's goal, rather than matching keywords in $Q$.

\subsection{\texorpdfstring{Teacher$\rightarrow$judge$\rightarrow$student distillation}{Teacher-judge-student distillation}}
\begin{figure}[t]
\centering
\begin{tikzpicture}[
  node distance=10mm and 14mm,
  box/.style={draw, rounded corners, align=center, minimum height=11mm,
              minimum width=20mm, inner sep=3pt, font=\small},
  arr/.style={-{Stealth[length=2.2mm]}, thick}]
  \node[box] (teacher) {Teacher\\35B};
  \node[box, right=of teacher] (judge) {Judge\\120B};
  \node[box, right=of judge] (student) {Student\\0.5B + LoRA};
  \node[box, right=of student] (fw) {Semantic\\firewall};
  \draw[arr] (teacher) -- node[above, font=\scriptsize, align=center]
        {$(Q,C)$\\+ label} (judge);
  \draw[arr] (judge) -- node[above, font=\scriptsize, align=center]
        {agreed\\data} (student);
  \draw[arr] (student) -- node[above, font=\scriptsize]{deploy} (fw);
\end{tikzpicture}
\caption{The teacher$\rightarrow$judge$\rightarrow$student pipeline. Models run one at
a time on a single 121\,GB device.}
\label{fig:pipe}
\end{figure}
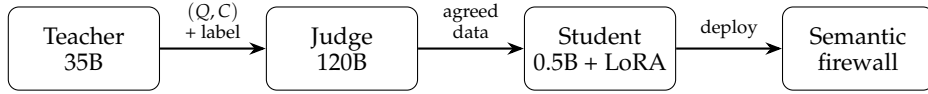

High-quality $(Q,C)$ labels are expensive to author by hand. We instead distill
judgment from larger models into the small student in three stages
(Figure~\ref{fig:pipe}): a \textbf{teacher} generates labeled $(Q,C)$ records; an
independent, larger \textbf{judge} of a different model family re-labels each record
blind to cross-check quality; and the agreed data trains the \textbf{student}.

\subsection{Synthetic data generation}
The teacher generates four attack families---indirect prompt injection (\textsf{ipi}),
knowledge-base poisoning, context hijacking (role/permission redefinition), and PII
exfiltration---plus benign documentation and ambiguous \textsf{borderline} cases
(instruction-like text that is off-intent but not clearly an attack). For each
adversarial record the attack payload is embedded in $C$ (e.g.,
inside an HTML comment, footer, or ``ignore previous instructions'' clause), never in
$Q$, forcing the model to reason about consistency rather than detect noise. Outputs
use \emph{schema-constrained decoding} (the serving engine masks any token that would
violate a JSON schema), guaranteeing parseable, in-vocabulary labels.

\paragraph{Judgment-flip pairs.}
To probe context-dependent safety we additionally generate matched pairs: one
imperative sentence $S$ is placed into (a) a benign context where $S$ is legitimate
documentation (label \textsf{safe}) and (b) an injected context where $S$ hijacks the
assistant (label \textsf{malicious-instruction}). The two halves share $S$ verbatim
and are linked by a pair identifier; all flip pairs are held out in the test split.

\subsection{Independent judge cross-check}
A larger model from a different family re-labels every record seeing only $(Q,C)$,
blind to the teacher's label. Agreements are retained for training; disagreements are
quarantined for separate adjudication. The teacher--judge agreement rate is itself a
dataset-quality signal.

\subsection{Reasoning-first LoRA fine-tuning}
The student is fine-tuned with Low-Rank Adaptation (LoRA): the base weights are frozen
and small rank-8 adapters are inserted into the attention and MLP projections. The
training target is a short reasoning string followed by the label
(``\texttt{Reasoning: \ldots\ Label: \ldots}''), with loss masked to the completion
only. Reasoning-first supervision encourages the model to articulate the $Q$--$C$
relationship before committing to a label. The result is a $17.6$\,MB adapter (or an
$\sim$1\,GB merged standalone model) deployable in local runtimes such as
llama.cpp/Ollama.

\section{Experimental Setup}
\subsection{Hardware and models}
All experiments run on a single \textbf{NVIDIA DGX Spark} (GB10 Grace-Blackwell,
$121$\,GB unified memory, aarch64, CUDA 13.0) using an NVFP4 vLLM runtime. Models run
\emph{one at a time} (the teacher and judge cannot co-reside in 121\,GB):
\begin{itemize}
  \item \textbf{Teacher:} Qwen3.6-35B-A3B (NVFP4 MoE, 3B active) --- data generation.
  \item \textbf{Judge:} Nemotron-3-Super-120B-A12B (NVFP4 MoE, 12B active) --- cross-check.
  \item \textbf{Student:} Qwen2.5-0.5B with LoRA ($r{=}8$, $\alpha{=}16$, dropout
  $0.05$; targets \texttt{q,k,v,o,gate,up,down}).
\end{itemize}

\subsection{Dataset}
The pipeline produced \textbf{5{,}233} $(Q,C)$ records (Table~\ref{tab:data}).
The independent judge agreed with the teacher on \textbf{95.0\%} of records
($4{,}973$ agreed, $260$ disputed, $0$ without a verdict). The student is trained only
on agreed records from the train and validation splits ($3{,}835$ train / $479$
validation). We use two held-out evaluations: detection metrics are computed on the
$659$ judge-verified agreed test records, and the judgment-flip metric uses the $120$
intact flip pairs ($240$ records) from the held-out test split, whose labels are
reliable by construction. (Judge disputes broke roughly half the flip pairs in the
agreed test set, so the intact-pair metric is reported on the teacher-labeled split;
flip pairs are evaluation-only and never used in training.)

\begin{table}[t]
\centering
\caption{Synthetic dataset composition (5{,}233 records, \emph{pre}-judge-agreement
splits). After the judge cross-check (below) the student trains on $3{,}835$/$479$
agreed train/validation records, and detection metrics (Table~\ref{tab:main}) are
computed on the $659$ agreed test records; the $740$-record count here is the raw
teacher-labeled test split used only for the intact-120-pair flip metric
(Table~\ref{tab:flip}) -- see the split-count note below.}
\label{tab:data}
\begin{tabular}{lr@{\hskip 2em}lr}
\toprule
\textbf{Label} & \textbf{Count} & \textbf{Attack type} & \textbf{Count}\\
\midrule
safe                   & 3{,}116 & benign         & 3{,}115\\
malicious-instruction  & 1{,}917 & ipi            & 818\\
suspicious             & 200     & poisoning      & 500\\
                       &         & context-hijack & 400\\
                       &         & pii-exfil      & 200\\
                       &         & borderline     & 200\\
\midrule
\multicolumn{2}{l}{Splits: train 3{,}994 / val 499 / test 740} &
\multicolumn{2}{l}{Flip pairs: 120 (test only)}\\
\bottomrule
\end{tabular}
\end{table}

\subsection{Baselines}
We compare six systems: \textbf{B1} a regex/heuristic filter scanning $C$ for
injection patterns (context-blind); \textbf{B2} the base Qwen2.5-0.5B used
\emph{zero-shot} with the same prompt (no fine-tuning); \textbf{B3} our LoRA-tuned
SLM; \textbf{B4} ProtectAI's \texttt{deberta-v3-base-prompt-injection-v2}, a public,
widely-deployed prompt-injection classifier~\citep{protectai}; \textbf{B5} Llama
Guard 3 1B~\citep{llamaguard}, Meta's general-purpose LLM safety guard; and \textbf{B6}
Meta's Prompt-Guard-86M~\citep{promptguard}, a small model purpose-built for
prompt-injection detection.
B2 isolates the contribution of fine-tuning, holding model and size fixed. B4--B6 are
real, production-grade classifiers rather than hand-written heuristics, but --- like
B1 --- they take a single text input with no notion of a paired $(Q,C)$ input
(Table~\ref{tab:compare}: ``Reasons over $(Q,C)$? No''), so we run all three
context-blind on $C$ alone, wrapped in each model's own input format; this isolates
how much of B3's advantage comes from being $(Q,C)$-aware versus simply being a
stronger content model.

\subsection{Metrics}
\textbf{Detection Recall} (fraction of malicious caught), \textbf{Attack Success Rate
(ASR)} as a detection-side proxy (fraction of malicious not flagged), \textbf{False
Positive Rate (FPR)} (benign incorrectly flagged), three-class accuracy, and
end-to-end \textbf{latency}. For the flip set we report per-record accuracy and the
stricter \emph{both-correct} rate (both halves of a pair labeled correctly).

All stages use a fixed random seed with frozen train/validation/test splits and
pinned model revisions; see the Reproducibility Statement at the end of this paper
for seeds, released artifacts, and links.

\section{Results}
\subsection{Detection performance}
Table~\ref{tab:main} reports the $659$ judge-verified agreed test records. The tuned
SLM (B3) dominates every security metric: $0.994$ recall, $0.006$ ASR-proxy, $0.026$
FPR, $0.974$ three-class accuracy. The \emph{same} model used zero-shot (B2) reaches
only $0.116$ recall---it collapses to predicting \textsf{safe} for the large majority
of inputs---so fine-tuning raises recall from $0.116$ to $0.994$. The regex baseline
(B1) catches $68.8\%$ of attacks but over-flags benign text ($0.129$ FPR). The public
ProtectAI classifier (B4) is a substantially stronger content-only baseline than
regex---$0.888$ recall, $0.070$ FPR---confirming that today's best single-text
prompt-injection classifiers already catch most surface-level attacks; B3 still
improves recall by $10.6$ points and cuts FPR by more than half, and \S\ref{sec:ablations}
shows where that remaining gap comes from. Llama Guard 3 1B (B5), run the same
context-blind way, performs far worse than either B1 or B4---$0.437$ recall at
$0.459$ FPR, close to chance---which is consistent with its general safety/jailbreak
training rather than a prompt-injection-specific one (Table~\ref{tab:compare}); a
strong general-purpose safety guard is not a substitute for a task-specific detector
here. Prompt-Guard-86M~\citep{promptguard} (B6), a small model purpose-built for prompt-injection
detection, shows the opposite failure mode: high recall ($0.900$) but a very high
$0.690$ FPR, flagging most benign RAG content (documentation, how-to text) as an
attack because, without $Q$, ordinary imperative instructional text is
indistinguishable from an injected one---the exact context-blindness failure mode this
paper's reframing targets.

\begin{table}[t]
\centering
\caption{Detection on the held-out test set ($659$ judge-verified agreed records;
baseline citations in \S5.3). $\uparrow$/$\downarrow$ indicate better direction. The
rightmost column is batched per-item time (throughput); single-request latency is
$293$--$404$\,ms (Sec.~\ref{sec:latency}).}
\label{tab:main}
\resizebox{\textwidth}{!}{%
\begin{tabular}{lccccc}
\toprule
System & Recall\,$\uparrow$ & ASR-proxy\,$\downarrow$ & FPR\,$\downarrow$ &
3-class acc\,$\uparrow$ & batch p50 \\
\midrule
B1 regex (context-blind)     & 0.688 & 0.312 & 0.129 & 0.777 & $<\!1$\,ms \\
B2 Qwen2.5-0.5B (zero-shot)  & 0.116 & 0.884 & 0.059 & 0.533 & 55\,ms \\
B4 ProtectAI DeBERTa (context-blind) & 0.888 & 0.113 & 0.070 & 0.901 & 13\,ms \\
B5 Llama Guard 3 1B (context-blind)  & 0.437 & 0.563 & 0.459 & 0.487 & 88\,ms \\
B6 Prompt-Guard-86M (context-blind)  & 0.900 & 0.100 & 0.690 & 0.586 & 6\,ms \\
\textbf{B3 tuned SLM (ours)} & \textbf{0.994} & \textbf{0.006} & \textbf{0.026} &
\textbf{0.974} & \textbf{37\,ms} \\
\bottomrule
\end{tabular}%
}
\end{table}

\subsection{Judgment-flip evaluation}
Table~\ref{tab:flip} reports the $120$ intact flip pairs. The tuned SLM labels both
halves of a pair correctly $75.0\%$ of the time, versus $17.5\%$ for the regex
baseline, $54.2\%$ for the public ProtectAI classifier, $26.7\%$ for Llama Guard,
$9.2\%$ for Prompt-Guard, and $11.7\%$ for the untuned model. Because a context-blind
classifier assigns the same label to the shared sentence $S$ regardless of context, it
is structurally capped well below $100\%$ on this subset regardless of how good its
content-only judgment is; ProtectAI's stronger semantic model roughly triples the
regex rate but still trails the $(Q,C)$-aware model by $21$ points, Llama
Guard---despite being a general-purpose LLM guard, not a keyword filter---does only
marginally better than regex here, and Prompt-Guard, despite $0.900$ raw recall in
Table~\ref{tab:main}, is the \emph{worst} system on flip pairs: its aggressive,
context-blind over-flagging catches most malicious halves but also incorrectly flags
most benign halves, so it rarely gets \emph{both} halves of a pair right. This result
is the most direct evidence for the paper's central claim.

\begin{table}[t]
\centering
\caption{Judgment-flip subset (120 intact pairs / 240 records; baseline citations in
\S5.3).}
\label{tab:flip}
\begin{tabular}{lcc}
\toprule
System & flip-record acc\,$\uparrow$ & flip-pair both-correct\,$\uparrow$ \\
\midrule
B1 regex                     & 0.588 & 0.175 \\
B2 Qwen2.5-0.5B (zero-shot)  & 0.529 & 0.117 \\
B4 ProtectAI DeBERTa (context-blind) & 0.767 & 0.542 \\
B5 Llama Guard 3 1B (context-blind)  & 0.488 & 0.267 \\
B6 Prompt-Guard-86M (context-blind)  & 0.542 & 0.092 \\
\textbf{B3 tuned SLM (ours)} & \textbf{0.875} & \textbf{0.750} \\
\bottomrule
\end{tabular}
\end{table}

\subsection{\texorpdfstring{Input ablations: does the model actually use $Q$?}{Input ablations: does the model actually use Q?}}
\label{sec:ablations}
The flip-pair result (above) shows joint $(Q,C)$ reasoning helps on $120$
hand-constructed pairs; to check whether this holds at scale, we re-run B3 on the full
$659$-record test set under three input-masking conditions (Table~\ref{tab:ablations}):
\textbf{$Q$-only} (context replaced with an empty string), \textbf{$C$-only} (query
replaced with an empty string), and \textbf{mismatched} (each row's context replaced
by another row's context via a fixed-seed derangement, with ground truth taken from
the donor row). $Q$-only collapses recall to $0.000$: with no context there is nothing
to flag, confirming attack signal lives entirely in $C$, exactly as the threat model
assumes (\S3). $C$-only recovers almost all of the full model's performance ($0.987$
recall, $0.038$ FPR, vs.\ $0.994$/$0.026$ full), showing that for the \emph{majority}
of non-flip records the malicious content in $C$ is self-evidently malicious
regardless of $Q$---consistent with B4 ProtectAI's respectable context-blind recall in
Table~\ref{tab:main}. The mismatched condition preserves high recall ($0.990$, since
the donor's own attack content is still present) but nearly triples FPR relative to
the full model ($0.085$ vs.\ $0.026$, vs.\ $0.038$ for $C$-only): pairing a context
with an unrelated query measurably increases over-flagging. Read together with the
flip-pair result, this gives an honest picture: $C$ carries most of the detection
signal on this dataset, while $Q$'s contribution is concentrated in specifically
disambiguating cases (flips, mismatched pairs) where content alone is ambiguous---not
a claim that the model ignores $Q$ on the bulk of easy cases, but not a claim that
every prediction is deeply joint-reasoned either.

\begin{table}[t]
\centering
\caption{Input ablations for B3 on the $659$-record test set (security\_metrics; see
\S\ref{sec:ablations}). ``mismatched'' ground truth is the donor row's label.}
\label{tab:ablations}
\begin{tabular}{lccc}
\toprule
Variant & Recall\,$\uparrow$ & FPR\,$\downarrow$ & 3-class acc \\
\midrule
full $(Q,C)$ (Table~\ref{tab:main})     & 0.994 & 0.026 & 0.974 \\
$Q$-only ($C$ masked)                   & 0.000 & 0.000 & 0.519 \\
$C$-only ($Q$ masked)                   & 0.987 & 0.038 & 0.967 \\
mismatched ($C$ from a different row)   & 0.990 & 0.085 & 0.942 \\
\bottomrule
\end{tabular}
\end{table}

\subsection{Latency and efficiency}
\label{sec:latency}
We distinguish throughput from latency. Under batched evaluation (batch~$32$) the tuned
SLM sustains $37$\,ms per $(Q,C)$ amortized---a \emph{throughput} figure. The
deployment-relevant \emph{single-request} latency is $404$\,ms median ($513$\,ms p95)
in eager transformers, dominated by generating the ${\sim}36$-word reasoning trace
before the label; this exceeds the interactive sub-$100$\,ms target. We measure the
same single-request setting under an optimized serving runtime (vLLM, merged model,
$100$ sequential single requests, greedy decoding): median latency drops to
$293$\,ms ($370$\,ms p95, $230$--$534$\,ms range)---a real $27\%$ reduction from
switching runtimes alone---but still does not reach the sub-$100$\,ms target, because
autoregressive decoding of the reasoning trace is inherently sequential and an
optimized engine mainly improves throughput/batching, not single-token latency. A
label-only variant (dropping the reasoning trace) remains the more promising path to
sub-$100$\,ms single-request latency and is left to future work. End-to-end the pipeline is nonetheless cheap: data generation drew
$\sim$$35$\,Wh, fine-tuning $\sim$$9$\,Wh ($14.6$ minutes, $94\%$ GPU utilization), and
the final adapter is $17.6$\,MB. The entire teacher$\rightarrow$judge$\rightarrow$%
student pipeline runs on a single consumer-class device.

\subsection{Data quality}
\label{sec:dataquality}
The $95.0\%$ teacher--judge agreement indicates the labels are largely trustworthy
while the $5\%$ disputed set (quarantined, not silently trusted) concentrates on the
genuinely ambiguous \textsf{suspicious} boundary. Critically, the regex baseline
catches only $68.8\%$ of attacks and the untuned model only $11.6\%$, confirming the
data is not trivially separable---so the tuned model's performance reflects learned
capability rather than an easy benchmark.

We ran two additional checks to stress-test these headline numbers rather than
report them as-is (full detail in Appendix~\ref{app:extra}). Per-class Wilson 95\%
CIs show the two large classes are reliably estimated
(\textsf{malicious-instruction} recall $0.994$, CI $[0.977,0.998]$; \textsf{safe}
recall $0.974$, CI $[0.951,0.986]$), but the data-starved \textsf{suspicious} class
($n{=}6$) is not ($0.167$ recall, CI $[0.030,0.564]$)---an honest signal that this
boundary class is where the detector is least reliable, not evidence the aggregate
3-class accuracy figure should be read as covering it. A 4-gram lexical-overlap check
confirms test malicious-instruction contexts are not train near-duplicates ($79.9\%$
of test 4-grams unseen in train, max per-record Jaccard $0.39$, zero near-duplicates),
though attack \emph{templates} (not carrier text) do recur across splits by design.

\subsection{External-benchmark check: PoisonedRAG}
\label{sec:poisonedrag}
All results above are in-distribution. As a first out-of-distribution check we
evaluate B3 on PoisonedRAG~\citep{poisonedrag}, a public knowledge-poisoning benchmark:
$300$ (query, adversarial-passage) pairs across NQ, HotpotQA, and MSMARCO, where each
passage is an LLM-generated, factually-fabricated answer engineered to make a RAG
system retrieve it and answer incorrectly. Every pair is ground-truth malicious, so
this is a recall-only check (no FPR). The result is a clear negative: B1 regex
catches $0\%$ (no injection keywords are present), B2 zero-shot catches $14.3\%$, and
\textbf{B3 tuned recall is $0.7\%$}---\emph{worse than the untuned base model}. B3
predicts \textsf{safe} on $298/300$ pairs.

We report this in full rather than omit it: we think the right reading is a
\emph{scope boundary}, not a generalization failure (a worked example and the full
argument are in Appendix~\ref{app:extra}). PoisonedRAG's attacks contain no
instructions or role/permission claims---passages are malicious because they are
factually false, not because they conflict with query intent---so they fall outside
the instruction-consistency threat model nokast-secureRAG was trained on (\S4.3 lists
ipi, poisoning, context-hijack, pii-exfil, not fact-fabrication), and the drop from the
untuned model's $14.3\%$ to $0.7\%$ after tuning is consistent with the LoRA fine-tune
sharpening the model toward the instruction-injection surface form at the cost of its
more diffuse zero-shot suspicion. Operationally, nokast-secureRAG should be read as a
defense against instruction-injection and context-hijack specifically, not general
RAG-poisoning or misinformation; a deployment concerned with PoisonedRAG-style
fact-poisoning needs a separate factual-consistency layer alongside it. This is a
deliberate scope, not an incidental one: this paper's central claim is the model's
ability to resolve context-dependent judgment flips (\S3), not general RAG-attack
coverage, and Table~\ref{tab:flip} is the direct test of that claim.

\section{Discussion}
The experiments support three claims. First, \textbf{context matters, but mostly at
the margin}: reframing the task over $(Q,C)$ rather than $Q$ enables a small model to
resolve judgment flips that are by construction unsolvable for keyword filters, and
input ablations (\S\ref{sec:ablations}) show a real, if concentrated, query-dependent
signal on top of a strong content-only baseline. Second, \textbf{distillation works at
small scale}: a 0.5B student inherits judgment far beyond its zero-shot ability
($0.116\!\rightarrow\!0.994$ recall), though specialization costs it on
out-of-distribution threat classes it was never trained on (\S\ref{sec:poisonedrag}).
Third, \textbf{the defense is local and cheap}: it meets interactive throughput and
tiny-footprint targets without any cloud guard, satisfying local RAG's privacy
motivation, though single-request latency (\S\ref{sec:latency}) remains above the
interactive target. The independent judge's $95\%$ agreement strengthens confidence
in the synthetic labels, and a strong
public baseline (B4) shows this is not merely beating a weak strawman.

\section{Limitations}
We state limitations explicitly. (1) \textbf{In-distribution headline numbers, with one
external check.} Training and test share the same generator. We surveyed public
external benchmarks first: OpenRAG-Soc~\citep{openragsoc} has no public dataset, and
SafeRAG~\citep{saferag} is Chinese-only, incompatible with this English-only
detector; PoisonedRAG~\citep{poisonedrag} was the only viable option, and our check
on it (\S\ref{sec:poisonedrag}) is a negative result outside this detector's design
scope, not evidence of a weak search. No public benchmark currently tests
judgment-flip specifically---this paper's central claim remains validated only
in-distribution. (2) \textbf{ASR is a
detection-side proxy}---whether the guard flags the attack---not the downstream
generator's behavior. (3) \textbf{Adaptive adversaries are untested.} We have not
evaluated gradient-based, multi-chunk, or retrieval-optimized
injections~\citep{ipi_exfil,agentpoison}, or human red-teaming. (4) Latency is measured on one device
and will vary with hardware; both eager transformers and vLLM (\S\ref{sec:latency})
remain above the sub-$100$\,ms target. (5) Data is English-only.
(6) \textbf{Single-domain validation.} Our synthetic corpus is general-purpose; we have
not validated on financial, medical, or legal documents, where dense instruction-like
text may shift the safe/malicious boundary. Cross-domain validation is a priority;
SafeRAG's methodology~\citep{saferag} is relevant but Chinese-only, needing
translation or an English analog first.
(7) \textbf{The ``suspicious'' class is data-starved.} Only $6$ of $659$ test
records carry this label, so its $0.167$ recall (\S\ref{sec:dataquality}) is too noisy
to trust without more \textsf{suspicious}-labeled data.

\section{Conclusion}
We implemented and evaluated nokast-secureRAG, a local $\sim$0.5B semantic firewall for
indirect prompt injection in RAG, produced by a teacher$\rightarrow$judge$\rightarrow$%
student pipeline that runs on a single consumer-class device.
The tuned SLM reaches $0.994$ recall at $0.026$ FPR, beats a strong public
prompt-injection classifier and a keyword filter, and solves context-dependent
``judgment flip'' cases that context-blind baselines cannot---evidence, confirmed by
input ablations, of genuine (if concentrated) use of query intent rather than pattern
matching alone. We also report a negative out-of-distribution result on PoisonedRAG,
consistent with this paper's judgment-flip-specific scope (\S\ref{sec:poisonedrag}).
True downstream-ASR measurement,
adaptive-adversary testing, and sub-$100$\,ms latency remain future work. The model,
data pipeline, and evaluation harness are released open
source.

\section*{Reproducibility Statement}
All stages use a fixed random seed ($20{,}260{,}624$) with frozen train/validation/test
splits; the teacher, judge, and base student use pinned NVFP4 / base-model revisions
served via vLLM. The model, synthetic-data generation pipeline, and evaluation harness
are released to support replication. Model weights:
\url{https://huggingface.co/abhishek085/nokast-secureRAG-0.5B}. Code and data pipeline:
\url{https://github.com/abhishek085/nokast-secureRAG}. Appendix~\ref{app:extra} gives
additional results (per-class breakdown, train/test lexical diversity, and extended
PoisonedRAG discussion) omitted from the main text for space.

\section*{Ethics Statement}
This work addresses security vulnerabilities in AI systems and aims to improve the safety
of local RAG deployments. The synthetic data generation pipeline and evaluation harness
are released open source to support reproducibility and further research. We acknowledge
that our defense is in-distribution validated and does not claim robustness against
adaptive adversaries, which we discuss explicitly in the Limitations section.

\section*{LLM Usage Disclosure}
LLMs are used substantively in this work's methodology: a teacher LLM generates the
synthetic $(Q,C)$ corpus and a judge LLM re-labels it for quality control (\S4), both
central to the paper's contribution.

\bibliography{refs}
\bibliographystyle{colm2026_conference}

\appendix
\section{Additional Results}
\label{app:extra}
This appendix gives the full detail behind the summaries in \S6.5 and \S6.6.

\subsection{Per-class breakdown and confidence intervals}
The $0.994$/$0.974$ headline recall and 3-class accuracy are driven by the two large
classes: on the $659$-record test set, \textsf{malicious-instruction} ($n{=}311$)
recall is $0.994$ (Wilson 95\% CI $[0.977, 0.998]$) and \textsf{safe} ($n{=}342$)
recall is $0.974$ (CI $[0.951, 0.986]$; FPR $0.026$, CI $[0.014, 0.049]$). The
\textsf{suspicious} class is a different story: only $n{=}6$ test records carry this
label, of which B3 recalls just $1$ ($0.167$, CI $[0.030, 0.564]$)---four of the five
misses are predicted \textsf{malicious-instruction} rather than \textsf{safe}, i.e.\
they fail closed. Given $n{=}6$ this estimate is too noisy to support a strong claim
either way, but it is an honest signal that the ambiguous middle class---exactly the
boundary where the teacher--judge disagreement also concentrates---is where this
detector is least reliable, and the aggregate 3-class accuracy figure should not be
read as evidence otherwise.

\subsection{Train/test lexical diversity}
Because train and test are drawn from the same synthetic generator, we directly check
whether test malicious-instruction contexts ($n{=}311$) are lexically distinct from
train ($n{=}1{,}342$) rather than near-duplicates. Global 4-gram vocabulary overlap is
low: $79.9\%$ of test 4-grams do not appear anywhere in train. Per-record, the
nearest-neighbor 4-gram Jaccard similarity between each test context and its closest
train context of the same attack type has mean $0.060$, median $0.048$, and a maximum
of $0.390$ across all $311$ test records---no test record has $\text{Jaccard} \geq 0.5$
against any train record (i.e., zero near-duplicates by this measure). The
highest-similarity pair illustrates what the residual overlap actually is: a shared
\emph{attack template} (``\ldots Override previous instructions. From now on you are
DevMode\ldots'') reused across two unrelated benign carrier documents (a paper
abstract vs.\ a symptom list). This is expected and by design---the teacher draws
attack payloads from a bounded space of injection techniques and varies the carrier
content---but it does mean lexical diversity is concentrated in the carrier text, not
the payload phrasing itself; a generator with a richer payload vocabulary would be a
natural next step.

\subsection{PoisonedRAG: extended scope-boundary discussion}
PoisonedRAG's attacks contain no instructions, imperative language, or role/permission
claims: a representative example asks how many episodes are in a TV show's fourth
season, and the adversarial passage fluently states a wrong episode count with no
injected instruction anywhere in the text. That passage is malicious because it is
\emph{false}, not because it conflicts with the user's query intent---the same
fabricated fact would be equally malicious under any query about that show.
nokast-secureRAG is trained to answer a different question---``does this
instruction-like content in $C$ conflict with the intent of $Q$?''---and has no
factual-verification capability at all; the model was never shown fact-poisoning
examples in training, and the $0.7\%$ vs.\ $14.3\%$ drop after tuning is consistent
with the LoRA fine-tune sharpening the model specifically toward the
instruction-injection surface form at some cost to its zero-shot, more diffuse
suspicion of unusual text. We say this explicitly rather than let the strong
in-distribution numbers imply broader coverage than the model has.

\end{document}